\RequirePackage{amsmath}
\documentclass[12pt]{iopart}
\pdfoutput=1 
\usepackage{graphicx,cite}
\usepackage{amsmath,amsfonts,amssymb}
\usepackage{bm,bbm}
\usepackage[urlcolor=blue]{hyperref}
\usepackage{graphicx}
\usepackage[T1]{fontenc}

\newcommand{\Elr}[1]{\left\langle #1\right\rangle}
\newcommand{\E}[1]{\langle #1\rangle}
\newcommand{\Es}[1]{\left\langle #1\right\rangle_{\rm s}}

\newcommand{\peq}{\rho_{\rm eq}}

\newcommand{\x}{ x}

\newcommand{\abs}[1]{\left\lvert #1 \right\rvert}
\newcommand{\norm}[1]{\left\lvert\left\lvert #1 \right\rvert\right\rvert}
\newcommand{\G}{G}

\newcommand{\Sigmas}{\Sigma_{\rm s}}
\newcommand{\Sigs}{\Sigma_{\rm s}}

\renewcommand{\P}[1]{\mathbb{P}\left[#1\right]}
\newcommand{\R}{\mathbb{R}}

\newcommand{\dT}{\delta \tilde T}

\newcommand{\dThc}{\delta \tilde T_{h,c}}

\newcommand{\rhot}{\rho(x,t)}
\newcommand{\rhos}{\rho_{\rm s}(x)}
\newcommand{\gat}{\Gamma(x,t)}
\newcommand{\gas}{\Gamma_{\rm s}(x)}
\newcommand{\nut}{\nu(x,t)}
\newcommand{\nus}{\nu_{\rm s}(x)}
\newcommand{\trhot}{\rho(x_t,t)}
\newcommand{\trhos}{\rho_{\rm s}(x_t)}
\newcommand{\tgat}{\Gamma(x_t,t)}
\newcommand{\tgas}{\Gamma_{\rm s}(x_t)}
\newcommand{\tnut}{\nu(x_t,t)}
\newcommand{\tnus}{\nu_{\rm s}(x_t)}

\usepackage{xcolor}
\definecolor{C0}{rgb}{0.12156862745098039, 0.4666666666666667, 0.7058823529411765}
\definecolor{C1}{rgb}{1.0, 0.4980392156862745, 0.054901960784313725}
\definecolor{C2}{rgb}{0.17254901960784313, 0.6274509803921569, 0.17254901960784313}
\definecolor{C3}{rgb}{0.8392156862745098, 0.15294117647058825, 0.1568627450980392}
\definecolor{C4}{rgb}{0.5803921568627451, 0.403921568627451, 0.7411764705882353}
\definecolor{C5}{rgb}{0.5490196078431373, 0.33725490196078434, 0.29411764705882354}
\definecolor{inferno0}{rgb}{0.087411, 0.044556, 0.224813}
\definecolor{inferno1}{rgb}{0.379001, 0.076253, 0.432719}
\definecolor{inferno2}{rgb}{0.658463, 0.178962, 0.372748}
\definecolor{inferno3}{rgb}{0.894305, 0.353399, 0.193584}
\definecolor{inferno4}{rgb}{0.987622, 0.64532 , 0.039886}

\colorlet{mylinkcolor}{blue!66!black!80}
\hypersetup{colorlinks = true, linkcolor = {mylinkcolor}, linkcolor = {mylinkcolor}, citecolor = {mylinkcolor}, urlcolor = {mylinkcolor}}
\definecolor{grey}{rgb}{0.6,0.6,.6}
\definecolor{darkgrey}{rgb}{0.4,0.4,.4}
\definecolor{darkgreen}{rgb}{0,0.4,0}
\definecolor{lightgreen}{rgb}{0,0.7,0}
\definecolor{darkred}{rgb}{0.5,0,0}

\newcommand{\blue}[1]{{\color{blue}#1}}

\renewcommand{\blue}[1]{{#1}}

\allowdisplaybreaks

\begin{document}
\title[Thermal relaxation asymmetry persists under inertial effects]{Thermal relaxation asymmetry persists under inertial effects}
\author{Cai Dieball$^1$ and Alja\v{z} Godec$^{2,1}$}
\address{$^1$ Mathematical bioPhysics Group, Max Planck Institute for Multidisciplinary Sciences, 37077 G\"ottingen, Germany}
\address{$^2$ Mathematical Physics and Stochastic Dynamics, Institute of Physics, University of Freiburg, 79104 Freiburg, Germany}
\ead{cai.dieball@mpinat.mpg.de}
\ead{agodec@physik.uni-freiburg.de}

\begin{abstract}
\blue{Relaxation phenomena far from thermal equilibrium feature a rich phenomenology, including the thermal relaxation asymmetry. The latter states that heating occurs faster than cooling in overdamped harmonic systems quantified in terms of excess free energy and entropy production. Upon isolating the relevant relaxational contribution to the entropy production, we here} algebraically prove the asymmetry in thermal relaxation in phase space in the entire range from overdamped dynamics to underdamped dynamics. We show that for the same setup as for overdamped dynamics \blue{(i.e., motion in harmonic potentials, also known as Ornstein-Uhlenbeck dynamics)}, even in the more general case of phase-space relaxation, i.e., 
underdamped dynamics, far-from-equilibrium heating is faster than cooling. 
The coupling of positions and velocities emerging in this generalization further underscores, in a striking manner, the intricate dynamics of such thermal relaxation processes that \emph{do not} pass through local equilibria. Investigating the overdamped limit, our generalized approach reveals, interestingly, that an excess free energy contribution from the velocity degrees of freedom  \emph{does not} trivially vanish in the overdamped limit, but is instead affected by the precise interpretation of temperature quenches in overdamped systems.
\end{abstract}

\section{Introduction}

According to the laws of thermodynamics and by common experience, systems in contact
with a thermal environment 
eventually settle to the temperature of their
surroundings in the process called \emph{thermal relaxation}
\cite{MazurGroot}. Relaxation close to
equilibrium may be explained by 
linear response theory which is in turn conceptually rooted in 
Onsager's regression hypothesis
\cite{onsager_1,onsager_2,Yokota}. That is, relaxation from a
\emph{weak} temperature quench is indistinguishable from the
decay from a 
corresponding thermal fluctuation occurring spontaneously at equilibrium
\cite{onsager_1,onsager_2,Yokota}. Analogous results were meanwhile 
formulated also for relaxation near non-equilibrium steady states
\cite{Baiesi_2013,Maes2011PRL,Polettini}. Far from (equilibrium and non-equilibrium) steady states, i.e.\ beyond the linear regime, however, the regression hypothesis and perturbative arguments fail, and manifestly non-equilibrium phenomena emerge (for a recent review see Ref.~\cite{Teza2026PR}). 

Important advances in understanding relaxation beyond
the linear regime include
second-order response \cite{second_1,second_2,second_3}, anomalous
diffusion \cite{Ralf,Igor},
hydrodynamic limits \cite{Bertini2004JSP,Bertini2015RMP}, barrier crossing in driven systems \cite{Maier1993PRE,Bouchet2016AHP}, memory effects
\cite{Sancho,Igor,Igor_2,Igor_3,Dima_2013,Dima_2019,Lapolla2019FP,Lapolla_2020,Haenggi_RMP,Netz},
far-from-equilibrium fluctuation-dissipation theorems
\cite{Cugliandolo1997PRL,Lippiello}, optimal heating/cooling protocols \cite{optimal}, 
anomalous relaxation phenomena a.k.a.\ the \emph{Mpemba effect} \cite{Lu2017MMpemba,Lasanta2017Mpemba,Baity2019MpembaSG,Kumar2020EMpemba,Carollo2021QMpemba,Kumar2022Anomalous,Klich,Oren_2022} (for a detailed recent account see \cite{Teza2026PR})
and its isothermal analog \cite{Deguenther2022EPL}, the 
Kovacs memory effect
\cite{Kovacs_protein,Militaru2021Kovacs},
and dynamical phase transitions 
\cite{Gaw,Cusp_3,Garrahan,Speck,Minimal,singu_2,Meibohm,Meibohm_2023,Kristian}. Further important and notable
advances further include transient
thermodynamic uncertainty relations 
\cite{Pietzonka2017PRE,Dechant2018JSMTE,Liu2020PRL,Koyuk2019PRL,Koyuk2020PRL,Dieball2023PRL},
speed limits
\cite{CSL_3,CSL_4,Ito_det,CSL_5,Saito},
and analyses of relaxation from the viewpoint of information
geometry \cite{Saito,CSL_5,Ibanez2024NP}. 

A particularly striking discovery was the asymmetry between
heating and cooling from thermodynamically equidistant (TE) temperature
quenches \cite{Lapolla2020PRL}: systems
with locally quadratic energy landscapes and microscopically
reversible dynamics heat up faster than they cool down.~Later works
further expanded on this result 
\cite{VanVu2021PRR,Manikandan2021PRR,Meibohm2021PRE}. The
asymmetry was
recently quantitatively confirmed by experiments \cite{Ibanez2024NP} and further proved for irreversible overdamped systems that are linearly driven out of equilibrium \cite{Dieball2023PRR}. The latter work also (quite spectacularly) underscored the non-existence of a temperature during the transients, by uncovering a rotation of level sets of probability densities in opposite directions during heating and cooling, respectively \cite{Dieball2023PRR}.    

From a physical point of view, the asymmetry is the result of the entropy production within
the system during heating being more efficient than the
heat dissipation
into the environment during cooling \cite{Lapolla2020PRL}. In turn, close to
equilibrium both mechanisms
become equivalent and consequently
symmetry is restored \cite{Lapolla2020PRL,Ibanez2024NP}. A deeper understanding
of the asymmetry was recently provided in terms of ``thermal kinematics''
\cite{Ibanez2024NP}, which shows that an early ``overshoot'' in relaxation velocity in the space of probability distributions during heating is sufficient to ensure faster heating.   

Notwithstanding, just as the persistence of the relaxation
asymmetry in overdamped systems driven irreversibly
into non-equilibrium steady states (NESS) was unexpected \cite{Dieball2023PRR},
the question whether the 
asymmetry emerges also in the presence of inertial effects in both, reversible as well as
linearly driven systems, remains elusive. Notably, recent Molecular Dynamics computer experiments with a Nos\'{e}–Hoover thermostatted system revealed signatures of the relaxation asymmetry \cite{Arabzadeh_2026}. 

Here, by substantially generalizing the algebraic proof in Ref.~\cite{Dieball2023PRR}, we prove the relaxation asymmetry also in the phase-space setting, i.e.\ by fully accounting for inertial effects during thermal relaxation. \blue{That is, we show that for underdamped dynamics in harmonic potentials, heating is faster than cooling quantified in terms of an excess free energy. Note that the statement is restricted to harmonic potentials, i.e., Ornstein-Uhlenbeck dynamics, as it breaks down for general potentials already in overdamped dynamics \cite{Lapolla2020PRL}.}
The coupling between positions and velocities
that emerges in this generalization further strikingly underscores that far-from-equilibrium thermal relaxation processes absolutely do \emph{not} pass through local equilibria.
Moreover, our generalized approach reveals that an excess free energy contribution from the velocity degrees of freedom does
\emph{not} trivially vanish in the overdamped limit, but is instead effected by the precise
interpretation of a temperature quench in overdamped systems, which so far was not considered. 

The work is structured as follows. \blue{In Sec.~\ref{Quant} we recapitulate stochastic thermodynamics of relaxation processes and the \emph{thermodynamically equidistant temperatures}. In particular, we target the quantification of relaxation for underdamped dynamics, which has so far received less attention than the overdamped case. This section lays the foundation that allows for precise statements for the results shown in the rest of the manuscript. In Sec.~\ref{Asym} we present the relaxation asymmetry in the most general form as a theorem. This is our main result and shows that the thermal relaxation asymmetry persists to underdamped dynamics, i.e., heating is faster than cooling in harmonic potentials even in the presence of inertial effects. In Sec.~\ref{limits} we reconcile two distinct viewpoints on the energetics of relaxation in the overdamped limit. Then, in Sec.~\ref{projections} we extend the proof of the asymmetry to dynamics projected onto position and velocity variables, respectively. This yields a further connection to the existing literature since the known overdamped cases solely address position variables. In Sec.~\ref{outlook} we provide a brief outlook on future directions. To improve the flow of reading, all mathematical proofs are presented in the final Sec.~\ref{sec:proof}.}

\section{Quantification of thermal relaxation}\label{Quant}
\subsection{The excess free energy}
Before we can address the properties of thermal relaxation, we first need to formalize how to quantify relaxation processes. 
There are a priori several ways to parametrize a relaxation in the space of probability densities $\rho(x,t)\overset{t\to\infty}\longrightarrow\rho_{\rm s}(x)$, where $\rho_{\rm s}(x)$ denotes the invariant density, and for the scope of this paper $\x=(r_1,r_2,\dots,r_d, v_1,v_2,\dots,v_d)^T$ is the phase-space vector of all positions and velocities. On the one hand, we want to quantify the time-dependent distance of $\rho(x,t)$ and $\rho_{\rm s}(x)$ with a mathematically established tool such as Wasserstein distances \cite{Jordan_1998,JORDAN1997265}, Kullback-Leibler divergences \cite{Kullback1951AMS} or information-geometric concepts \cite{Shiraishi2018PRL,CSL_5, Ibanez2024NP}. On the other hand, one should monitor thermodynamic observables such as entropy production, dissipated heat or free energy to give thermodynamic meaning to the mathematical description of the relaxation dynamics. 

A particular quantity that combined the notion of a mathematically consistent quantification with a strong thermodynamic interpretation was found in the excess free energy (EFE) \cite{Lebowitz1957AP,Vaikuntanathan2009EEL,VandenBroeck2010PRE,Lapolla2020PRL}
\begin{align}
    \mathcal D(t)
    =k_{\rm B}T\int\rmd\x\, \rho(x,t)\ln\left[\frac{\rho(x,t)}{\rho_{\rm s}(x)}\right]
    \,.\label{EFE}
\end{align}
Mathematically speaking, this involves the Kullback-Leibler divergence $\mathcal D(t)=k_BTD_{KL}[\rho(x,t)||\rho_{\rm s}(x)]$ that fulfills desired properties like $\mathcal D(t)\ge 0$ and $\mathcal D(t)=0$ if and only if $\rho_{\rm s}(x)=\rho(x,t)$ for (almost) all $\x$. At the same time, $\mathcal D(t)$ has strong thermodynamic meaning in terms of both, free energy considerations and entropy production, as we now explain in detail.

\subsection{Connection to excess free energy}
We here recall the established interpretation of the excess free energy \cite{Vaikuntanathan2009EEL,Lapolla2020PRL,Ibanez2024NP} and generalize it to underdamped dynamics.  
Note that the relaxation will always occur at one unique temperature $T$ since we assume the temperature to be quenched at $t=0$.
Consider first underdamped Langevin dynamics in $d$-dimensional space ($2d$-dimensional phase space) \cite{Risken1989}
\begin{align}
    \rmd r_t &=  v_t\rmd t\nonumber\\
     m\rmd v_t &=  F( r_t, v_t)\rmd t-\gamma  v_t\rmd t+\sqrt{2k_BT\gamma}\rmd W_t\,.\label{LE general}    
\end{align}
where $ m$ and $\gamma$ are  $d\times d$ matrices that are positive definite (i.e., symmetric with positive eigenvalues, such that $\sqrt{\gamma}$ is well-defined and unique up to orthogonal changes of basis, that can be absorbed in the rotation-invariant statistics of $W_t$) and $ F$ is the force vector (without the friction force). For now, assume that this dynamics approaches an equilibrium density $\rho_{\rm s}(x)=\peq(x)$, where the phase-space vector $\x=(r_1,r_2,\dots,r_d, v_1,v_2,\dots,v_d)^T$ contains both the position $ r=(r_1,r_2,\dots,r_d)$ and velocity vector $ v=(v_1,v_2,\dots,v_d)$. For conditions required for the existence of an invariant measure for phase-space diffusion see \cite{Kliemann_1987}.
Expressed in this phase-space notation, Eq.~\eqref{LE general} reads
\begin{align}
  dx_t&=\begin{bmatrix} v_t \\ m^{-1}F(r_t,v_t)-m^{-1}\gamma v_t\end{bmatrix} dt+\begin{bmatrix} 0 & 0\\ 0 & \sqrt{2k_{\rm B}T\gamma}   \end{bmatrix} dW_t
  \,.\label{LE phase space general}
\end{align}
For cases admitting physical interpretation, we assume that the equilibrium density is of the Maxwell-Boltzmann form (later derivations only demand that $\phi(x)$ is sufficiently confining, but for physical interpretations it is more useful to assume this form)
\begin{align}
    \peq(x)&=\exp[-\phi(x)/k_BT]/Z\nonumber\\
    \phi(x)&=\phi_r( r)+\frac12m v^2\,,
\end{align}
where $\phi_r( r)$ is a sufficiently confining spatial potential.

Following the steps of Ref.~\cite[Sec.~5.2]{Dieball_Thesis} (see also Ref.~\cite{Seifert2012RPP}), we have that $F_{\rm eq}=-k_BT\ln Z$ (for a derivation, consider Eq.~\eqref{t_free_energy} with $\rho(x,t)=\peq(x)$) is the equilibrium free energy and with the internal energy $U=\Es{\phi(x_t)}$ and Gibbs-Shannon entropy of the system $S_{\rm system}=-k_{\rm B}\int\rmd\x\peq(x)\ln[\peq(x)]$ we obtain for a time-dependent density $\rho(x,t)$ that the 
free energy $F=U-TS$ obeys
\begin{align}
    F(t)&=\E{\phi(x_t)}-TS_{\rm system}(t)\nonumber\\
    &=\int\rmd\x\, \rho(x,t)\phi(x)+k_{\rm B}T\int\rmd\x\, \rho(x,t)\ln[\rho(x,t)]\nonumber\\
    &=-k_{\rm B}T\int\rmd\x\, \rho(x,t)(\ln[\peq(x)]+\ln Z)+k_{\rm B}T\int\rmd\x\, \rho(x,t)\ln[\rho(x,t)]\nonumber\\
    &=F_{\rm eq}+k_{\rm B}T\int\rmd\x\, \rho(x,t)\ln\left[\frac{\rho(x,t)}{\peq(x)}\right]\nonumber\\
    &=F_{\rm eq}+\mathcal D(t)\,.\label{t_free_energy}
\end{align}
Thus, for relaxation towards equilibrium, $\mathcal D(t)$ measures the excess of free energy as compared to $F_{\rm eq}$.

\subsection{Connection to entropy production}
Exactly as for overdamped dynamics \cite[Sec.~5.2.2]{Dieball_Thesis}, the EFE at time $t$ corresponds to the total entropy production $S_{\rm total}=S_{\rm medium}+S_{\rm system}=Q/T+S_{\rm system}$ \cite{Seifert2012RPP,Seifert2025} of the process during the relaxation towards equilibrium (i.e., in the time interval $[t,\infty)$). Here, the entropy production in the medium at constant temperature $T$ corresponds to the released heat $TS_{\rm medium}=Q$.
Using that the latter is obtained from the gradient of the potential $Q=\E{\int_{\tau=t}^{\tau=\infty}-\nabla_\x\phi(x_\tau)\circ \rmd\phi(x_\tau)}$ during relaxation to equilibrium \cite{Seifert2012RPP}, we obtain (as for overdamped \cite{VandenBroeck2010PRE}, see Ref.~\cite[Sec.~5.2.2]{Dieball_Thesis} for the exact steps, and see Ref.~\cite[Eqs.~(A2), (A5)]{Lyu2024PRE} to confirm that all expressions directly generalize to underdamped dynamics via the choice $\phi(x)=\phi_{r}( r)+\frac12m v^2$)
\begin{align}
    \mathcal D(t) &= F(t)-F \nonumber\\
    &= \E{\phi(x_t)-\phi(x_\infty)}-T[S_{\rm system}(t)-S_{\rm system}(\infty)]\nonumber\\
    &= \E{\phi_r(r_t)-\phi_r(r_\infty)}+\frac12m\E{v_t^2-v_\infty^2}-T[S_{\rm system}(t)-S_{\rm system}(\infty)]\nonumber\\
    &=\Elr{\int_{\tau=t}^{\tau=\infty}-\nabla_r\phi_r(r_\tau)\cdot\rmd r_\tau}+\Elr{\int_{\tau=t}^{\tau=\infty}-mv\cdot\circ\rmd v_\tau}+T\Delta S_{\rm system}([t,\infty))\nonumber\\
    &=\Elr{\int_{\tau=t}^{\tau=\infty}-\nabla_\x\phi(x_\tau)\cdot\circ\rmd\x_\tau}+T\Delta S_{\rm system}([t,\infty))\nonumber\\
    &=T\Delta S_{\rm medium}([t,\infty))+T\Delta S_{\rm system}([t,\infty)) \nonumber\\
    &= T\Delta S_{\rm total}([t,\infty))\,.
\end{align}
Here, $\nabla_\x\equiv(\nabla_{ r},\nabla_{ v})^T$ and $\circ$ denotes the Stratonovich integral, which coincides with the standard (or any other convention like It\^o or anti-It\^o) integral in the (differentiable) $ r$-components. 

\subsection{Relaxation towards non-equilibrium steady states}
For relaxation towards non-equilibrium steady states (NESS) with invariant density $\rho_{\rm s}(x)$, the quantity $\mathcal D(t)$ as defined in Eq.~\eqref{EFE} is still the appropriate observable to quantify thermal relaxation \cite{Dieball2023PRR}. While the interpretation as free energies and total entropy production are lost for relaxation towards NESS, $\mathcal D(t)$ then corresponds to the \emph{non-adiabatic part of the entropy production} \cite{VandenBroeck2010PRE} (also see derivation in \cite[Sec.~5.2.2]{Dieball_Thesis}), and can be argued to be the appropriate observable for thermal relaxation, see Refs.~\cite{Dieball2023PRR} and \cite[Sec.~5.2.2]{Dieball_Thesis}. 

In the underdamped (phase-space) setting for relaxation towards NESS, the relaxatory part is still well described by the $\mathcal D(t)$. In Sec.~\ref{sec:proof_underdamped_entropy}, we show that the rate of total entropy production can be decomposed similarly to the overdamped case \cite{VandenBroeck2010PRE} as (see also Ref.~\cite{Spinney2012PRE} for discussions of further splittings of $S_{\rm tot}$)
\begin{align}
\partial_t S_{\rm tot}/k_{\rm B}=\frac{\partial_t\mathcal D(t)}{k_{\rm B}T}+\Elr{\tnus\gamma\tnus+2\begin{bmatrix} m^{-1}p_t \\  F(q_t)\end{bmatrix}\cdot\nabla\ln\trhos}\,.
\end{align}
Here, the third term does not appear in the overdamped case. However, since similar to the second term (which is in the case the overdamped the \textit{adiabatic} part \cite{VandenBroeck2010PRE}), it only involves $\rhot$ via the average $\E{\cdots}$, we decide to attribute this term to the adiabatic or \textit{housekeeping} part of the entropy production and (as in the overdamped case \cite{Dieball2023PRR}) only consider the $\mathcal D(t)$-part for the interpretation of relaxation. 

Note that the physical interpretation is still weaker than in the relaxation towards equilibrium since $\mathcal D(t)$ does \emph{not} contain the full entropy production and the interpretation as a free energy does not apply to driven systems. For convenience, we nevertheless generally refer to $\mathcal D(t)$ as the EFE.

\subsection{Thermodynamically equidistant initial conditions}

To compare heating and cooling dynamics at an ambient temperature $T$, consider temperatures $T_c<T<T_h$ and relaxation in a potential $\phi(x)$ starting from the Maxwell-Boltzmann initial condition $\rho_i(x,t=0)\propto\exp[-\phi(x)/k_BT_i]$ for $i=h,c$.
Note that the relaxation process solely occurs at the temperature $T$ since we assume the environment to be quenched from temperature $T_{c,h}$ to $T$ at time $t=0$. 
The temperatures $T_{c,h}$ are called \textit{thermodynamically equidistant (TE)} if $\mathcal D_c(0)=\mathcal D_h(0)$ where $\mathcal D_i(0)$ is the EFE for the process starting from $\rho_i(x,t=0)$. 

\section{Thermal relaxation asymmetry generalized to underdamped dynamics}\label{Asym}
\subsection{State of the art: Thermal relaxation asymmetry for overdamped dynamics}
Before generalizing, we briefly recall the established results on the thermal relaxation asymmetry for overdamped dynamics. 
It was theoretically predicted \cite{Lapolla2020PRL} and later experimentally verified \cite{Ibanez2024NP} that for \emph{overdamped} relaxation in a harmonic potential (also known as Ornstein-Uhlenbeck process (OUP)) starting from TE temperatures, we have $\mathcal D_c(t)<\mathcal D_h(t)$ for all $t>0$. Since we have $\lim_{t\to\infty}\mathcal D_i(t)=0$, this means that the heating process (i.e., the EFE $\mathcal D_c(t)$ starting from $T_c$) is faster than the cooing process, since the former is for any $t>0$ closer to the final value $\lim_{t\to\infty}\mathcal D_i(t)=0$. 

\subsection{Thermal relaxation asymmetry as a theorem}

We now present the main result of the work, 
%
formulated in the following theorem:
\medskip

\noindent \textit{\textbf{Theorem 1:} Let $W_t$ be the $n$-dimensional Wiener process and consider the $n$-dimensional SDE
\begin{align}
    \rmd x_t=-Ax_t\rmd t+\sigma\rmd W_t\,,\label{SDE in theorem}
\end{align}
with matrix $A\in\R^{n\times n}$ whose eigenvalues have exclusively
positive real parts, and a 
noise amplitude matrix $\sigma\in\R^{n\times n}$. Assume that the process in Eq.~\eqref{SDE in theorem} approaches an invariant Gaussian density with positive definite covariance matrix $\Sigs$. Then, for any pair of TE initial conditions (dictating zero-mean Gaussian initial conditions $\Sigma_{h,c}(0)=T_{h,c}\Sigs/T$) we have $\mathcal D_c(t)<\mathcal D_h(t)$ for all $t>0$.\\}
\medskip

The proof will be given later in Sec.~\ref{sec:proof_thm1}. For convenience, we will restate the claim in the theorem in terms of $\Delta\mathcal D(t) \equiv \mathcal D_h(t)-\mathcal D_c(t)\ge 0$. Technically, Theorem 1 contains both, overdamped ($x=r$, $n=d$) and underdamped dynamics ($x=(r,v)^T$, $n=2d$) approaching equilibrium or NESS (see examples below). The diffusion matrix is generally given by $D=\sigma\sigma^T/2$. We illustrate this theorem in the following for underdamped dynamics.

\subsection{Dynamics of the Ornstein-Uhlenbeck process}
Before we explain and understand the implications of the theorem (and the role of the assumptions) in more detail, we briefly outline the dynamics of the OUP as in Eq.~\eqref{SDE in theorem}. Given any Gaussian initial condition, the process is Gaussian at all times see, e.g., Refs.~\cite{Gardiner1985,Pavliotis2014TAM}. Thus, the probability density is characterized by the mean $\E{x_t}$ and covariance $\Sigma(t)\equiv\E{x_tx_t^T}-\E{x_t}\E{x_t^T}$, that obey the dynamical equations (see, e.g., \cite[Supplemental Material]{Dieball2023PRR} for a short derivation)
\begin{align}
    \rmd\E{x_t} &= -A\E{x_t}\rmd t\nonumber\\
    \rmd \Sigma(t) &= -A\Sigma(t)-\Sigma(t)A^T+2D\,.\label{dynamic Lyapunov}
\end{align}
If the system approaches a steady state $\rmd \Sigma(t)/\rmd t=0$, the steady-state covariance $\Sigs$ is given by 
\begin{align}
   A\Sigs + \Sigs A^T &= 2D\,,\label{static Lyapunov}
\end{align}
which allows expressing $\E{x_t}$ and $\Sigma(t)$ in terms of the initial values $\E{x_0}$ and $\Sigma(0)$ as (see, e.g., \cite[Supplemental Material]{Dieball2023PRR} for a short derivation)
\begin{align}
    \E{x_t} &= \rme^{-At}\E{x_0}\nonumber\\
    \Sigma(t) &= \Sigmas+\rme^{- At}[\Sigma(0)-\Sigmas]\rme^{- A^Tt}\,.\label{solution dynamic Lyapunov}
\end{align}
Note that for thermal relaxation considerations, Eq.~\eqref{static Lyapunov} implies that the temperature quench corresponds to dynamics at temperature $T$ starting from an initial covariance $\Sigma(0)_{h,c}=T_{h,c}\Sigs/T$ since $A$ is assumed to be independent of $T$ while $D\propto T$ (see also Ref.~\cite{Dieball2023PRR}).
Further note that the covariance dynamics in Eq.~\eqref{solution dynamic Lyapunov} further emphasize a reoccurring observation with anomalous thermal relaxation dynamics \cite{Dieball2023PRR,Ibanez2024NP}: The density $\rho(x,t)$ at intermediate times $t>0$ does not reflect states of temperature between $T_{c,h}$ and $T$. Instead, in general only $\Sigma(0)$ and $\Sigma(t\to\infty)$ are proportional to $\Sigs$ (and thus have $\rho(x,t)$ correspond to steady states of a defined temperature) and the intermediate $\rho(x,t)$ contain further contributions, in this case, e.g., reflecting transient $r$-$v$-correlations (other cases, e.g., in Refs.~\cite{Dieball2023PRR,Ibanez2024NP}).

\subsection{Assumptions in the theorem}
With the dynamical equations above, the assumptions made in the theorem can be mathematically and physically motivated and explained. First, the exclusively positive real parts of eigenvalues of $A$ ensure that $\rme^{-At}\to 0$ as $t\to\infty$ which is a prerequisite of convergence to a steady state. In terms of motion in a harmonic potential, this positivity corresponds to all (eigen-)directions of the potential being confining, although for underdamped dynamics, $A$ also contains the coupling of $r$ and $v$ such that the naive notion of a potential becomes problematic, see examples in the next (sub)section.

Next, note that we do \emph{not} make any assumptions on $\sigma$ in Eq.~\eqref{SDE in theorem}. Since $D=\sigma\sigma^T/2$, the diffusion matrix $D$ is always positive semi-definite (this is the case for any matrix of the form $MM^T$ with real $M$ because $(MM^T)^T=MM^T$ and for eigenvector and eigenvalue $MM^Tv=\lambda v$, we have $\lambda v^2=v^TMM^Tv=(M^Tv)^2\ge 0$ and $v^2>0$). For overdamped dynamics, we usually have to assume positive definiteness, while for underdamped dynamics, $D$ always has zero eigenvalues (see examples below). 
While a positive definite $D$ in Eq.~\eqref{static Lyapunov} implies a positive definite $\Sigs$, a positive semi-definite $D$ generally only implies that $\Sigs$ is positive semi-definite \cite{Bhatia2002}. Physically speaking, all degrees of freedom have to (at least indirectly) experience thermal noise to ensure equilibration, corresponding to a positive definite $\Sigs$. That means that noise acting only the velocities (i.e., semi-definite $D$) suffices to obtain a positive definite $\Sigs$ because the positions experience thermalization via their dependence on velocities. 

This idea of coupling the degrees of freedom via $A$ such that all experience thermalization for semi-definite $D$ is formalized as follows: Under the assumption that $A$ has exclusively eigenvalues with positive real parts, the definiteness of $\Sigs$ is equivalent to $\det(\int_0^t\rme^{-As}D\rme^{-A^Ts}\rmd s)>0$ for all $t>0$ \cite{Metafune_2,DaPrato2014}.
Alternatively, $\Sigs$ may be explicitly known and thus identified to be positive definite, e.g., via the equipartition theorem, see examples below.

\subsection{The underdamped OUP}
Phrasing Eq.~\eqref{SDE in theorem} (stochastic dynamics with linear drift) in terms of underdamped motion in $d$-dimensional space, i.e., $2d$-dimensional phase-space, we consider $x=(r,v)^T=(r_1,r_2,\dots,r_d, v_1,v_2,\dots,v_d)^T$ and write (with mass $m$ and friction $\gamma$ as introduced in Eq.~\eqref{LE phase space general})
\begin{align}
    \rmd r_t &= v_t\rmd t\nonumber\\
    \rmd v_t &= -m^{-1} A_rr_t\rmd t - m^{-1} (A_v+\gamma)v_t\rmd t + m^{-1}\sqrt{2k_BT\gamma}\rmd W_t\,.\label{underdamped SDE rv}
\end{align}
To compare with the form of Eq.~\eqref{SDE in theorem}, write Eq.~\eqref{underdamped SDE rv} in block matrix form (with $d\times d$ blocks) as
\begin{align}
    \rmd x_t &= -Ax_t\rmd t+\sigma\rmd W_t\nonumber\\
    A&=\begin{bmatrix}0&-1\\m^{-1} A_r&m^{-1} (A_v+\gamma)\end{bmatrix}\nonumber\\
    \sigma&=\begin{bmatrix}0&0\\0&m^{-1}\sqrt{2k_BT\gamma}\end{bmatrix}\,.\label{underdamped SDE x}
\end{align}
The notion of equilibrium underdamped motion in a harmonic (spatial) potential $\phi_r(r)$ is a special case of Eq.~\eqref{SDE in theorem} and corresponds to a scalar $m>0$ and symmetric $A_r$ with positive eigenvalues (such that $\phi_r(r)=r^TA_rr$ is confining) and $A_v=0$, see following example. 

For the underdamped motion we require, as mentioned above, that the eigenvalues of $A$ (i.e., not only of $A_r$) have positive real parts. We show in Sec.~\ref{sec:proof_positive_real_parts} that (as to be expected), for scalar $m$ and $\gamma$ and $A_r$ with positive \emph{real} eigenvalues, this is always the case. However, in the second example (see Sec.~\ref{sec:example_2}), we show that this stability (i.e., $A$ having exclusively positive real parts of eigenvalues) may break down for stable $A_r$ that have complex eigenvalues (but still positive real parts, i.e., the overdamped motion in this drift field would still be stable).

\subsection{The $1+1$-dimensional underdamped OUP}\label{sec:example_1}
In one-dimensional space, i.e., two-dimensional phase space $x=(r,v)^T$, Eq.~\eqref{underdamped SDE x} for underdamped motion in a harmonic energy potential $\phi_r(r)=ar^2/2$ becomes ($a,\gamma,m>0$)
\begin{align}
    \rmd x_t &= -Ax_t\rmd t+\sigma\rmd W_t\nonumber\\
    A&=\begin{bmatrix}0&-1\\a/m&\gamma/m\end{bmatrix}\nonumber\\
    \sigma&=\begin{bmatrix}0&0\\0&\sqrt{2k_BT\gamma}/m\end{bmatrix}\,.\label{underdamped SDE x 1d}
\end{align}
This dynamics asymptotically approaches the well-known equipartition steady-state covariance
\begin{align}
    \Sigs=k_BT\begin{bmatrix} \frac1a & 0 \\ 0& \frac1m \end{bmatrix}\,,
\end{align}
which can be verified via Eq.~\eqref{static Lyapunov}. Thus, $\Sigs$ is clearly positive definite and $\rho(x,t)$ approaches for $t\to\infty$ a zero-mean Gaussian with covariance $\Sigs$. Some exemplary trajectories and the excess free energies during heating and cooling for this example are shown in Fig.~\ref{fig:trajectories_and_EFE}a,b.

\subsection{Driven $2+2$-dimensional underdamped OUP}\label{sec:example_2}
To also address irreversibly driven dynamics (i.e., thermal relaxation into NESS), we need to consider at least two-dimensional space $x=(r,v)^T=(r_1,r_2,v_1,v_2)^T$. To this end, we consider a harmonic two-dimensional spatial potential $\phi_r(r)=a(r_1^2+r_2^2)/2$. In addition, we introduce a linear non-equilibrium drift in space via a parameter $\omega$ (for elaboration on this in the overdamped setting see Ref.~\cite{Dieball2023PRR}).
The equation of motion becomes ($a,\gamma,m>0$ and $\omega\in\R$)
\begin{align}
    \rmd x_t &= -Ax_t\rmd t+\sigma\rmd W_t\nonumber\\
    A&=\begin{bmatrix}0&0&-1&0\\0&0&0&-1\\a/m&\omega/m&\gamma/m&0\\-\omega/m&a/m&0&\gamma/m\end{bmatrix}\nonumber\\
    \sigma&=\begin{bmatrix}0&0&0&0\\0&0&0&0\\0&0&\sqrt{2k_BT\gamma}/m&0\\0&0&0&\sqrt{2k_BT\gamma}/m\end{bmatrix}\,.\label{underdamped SDE x 2d}
\end{align}
Note that the eigenvalues of $A_r\equiv\begin{bmatrix}a&\omega\\-\omega&a\end{bmatrix}$ have always positive real parts. However, in the case of complex eigenvalues of this spatial matrix, we did \emph{not} prove that this implies positive real parts of eigenvalues of $A$, see proof in Sec.~\ref{sec:proof_positive_real_parts}. By directly investigating the eigenvalues of $A$, we see that the real parts only remain positive for small/moderate driving. 

For example, if $a=m=\gamma=k_BT=1$ the real parts of all eigenvalues remain positive for $\abs{\omega}<1$. However, for $\abs{\omega}>1$, two of the eigenvalues have negative real parts. The latter signals coherent oscillations that destroy the confining character of the linear drift $A$, such that trajectories drift off to $\infty$ and an invariant density ceases to exist. Note that this does \emph{not} occur in overdamped dynamics due to the absence of coherence---they remain stable for all $\omega\in\R$ as seen from the positive real parts of the eigenvalues of $A_r$.

\begin{figure}
    \centering
    \includegraphics[width=.9\linewidth]{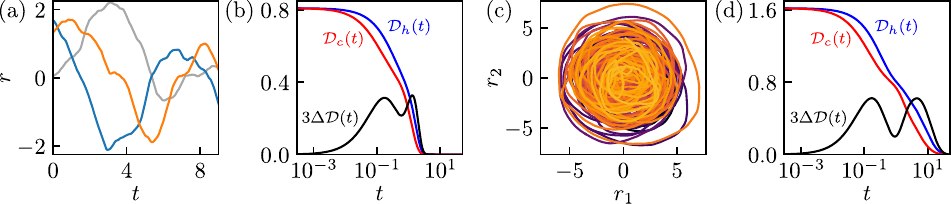}
    \caption{(a) Position component of three trajectories (generated via Euler algorithm with $\rmd t=0.005$) and (b) EFE for dynamics in Eq.~\eqref{underdamped SDE x 1d} with $a=m=\gamma=k_BT=1$ and TE temperatures $k_BT_c=0.2$ and $k_BT_h=2.86$.~(c) Trajectory of total length $t=1000$ ($\rmd t=0.002$) with time running from dark to bright and (d) EFE for dynamics in Eq.~\eqref{underdamped SDE x 2d} with $a=m=\gamma=k_BT=1$, $\omega=0.9$ and TE temperatures $k_BT_c=0.2$ and $k_BT_h=2.86$. The black line in (b) and (d) denotes $\Delta\mathcal D(t) \equiv \mathcal D_h(t)-\mathcal D_c(t)$ and is scaled up by a factor 3 for improved visibility.}
    \label{fig:trajectories_and_EFE}
\end{figure}
In Fig.~\ref{fig:trajectories_and_EFE}c we show a long trajectory for $\omega=0.9$ where close-to-coherent oscillations are visible. Note that the thermal relaxation asymmetry (Theorem 1) also applies to such coherent driven systems (as long as confinement is given), as shown in the EFE in Fig.~\ref{fig:trajectories_and_EFE}d.

\begin{figure}
    \centering
    \includegraphics[width=.9\linewidth]{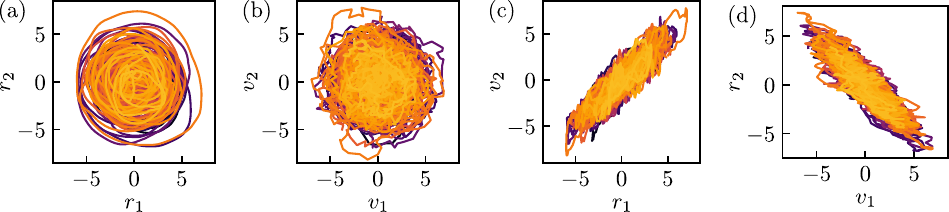}
    \caption{Trajectory from Fig.~\ref{fig:trajectories_and_EFE}c shown for several $r$- and $v$-components.}
    \label{fig:correlated_steady_state}
\end{figure}

To emphasize the difference between driven and equilibrium dynamics, and also the difference between underdamped and overdamped driven dynamics, we show in Fig.~\ref{fig:correlated_steady_state} the same trajectory as in Fig.~\ref{fig:trajectories_and_EFE}c, but different combinations of $r$- and $v$-components. This reveals that the rotations (in counterclockwise direction) induced by $\omega>0$ lead to steady-state correlations of $r$ and $v$. Note that this starkly differs from the Maxwell-Boltzmann distribution (which does \emph{not} apply since this is a NESS and \emph{not} equilibrium). Technically, since we have a long trajectory,  Fig.~\ref{fig:correlated_steady_state} represents the marginals of the Gaussian density with covariance according to the solution of the Lyapunov equation~\eqref{static Lyapunov} for $A$ and $D$ as in Eq.~\eqref{underdamped SDE x 2d} with $a=m=\gamma=k_BT=1$ and $\omega=0.9$, which is obtained numerically as
\begin{align}
    \Sigs &= \begin{bmatrix}5.26&0&0&4.74\\0&5.26&-4.74&0\\0&-4.74&5.26&0\\4.74&0&0&5.26\end{bmatrix}\,.
\end{align}

\section{The two overdamped limits}\label{limits}

Since the dynamics of the overdamped OUP are obtained as the limit $m/\gamma\to 0$ from the underdamped OUP (as, e.g., in Eq.~\eqref{underdamped SDE x 1d}) \cite{Risken1989,Gardiner1985}, one would assume that the thermal relaxation asymmetry in the overdamped case is contained as a special case of the underdamped one. While this is indeed the case, taking this limit surprisingly reveals an ambiguity (but \emph{not} a contradiction) in the overdamped energetics. \blue{Here, we outline and reconcile two ways of quantifying the energetics of in the overdamped limit.}

The key characteristic of the overdamped limit is that the timescale of velocity relaxation approaches $0$, i.e., the velocities instantly relax to a Gaussian distribution with variance $\propto T$. Recall that we consider the initial condition of the thermal relaxation process to correspond to the steady state at temperature $T_i$. Then the relaxation process happens at temperature $T$ and the position and velocity approach a variance $\propto T$ starting from $\propto T_i$. However, in the overdamped description, the velocities do not appear explicitly and are at temperature $T$ always assumed to be at variance $\propto T$. Therefore, the existing overdamped literature addressing the thermal relaxation asymmetry \cite{Lapolla2020PRL,Dieball2023PRR,Ibanez2024NP} does \textit{not} account for the free energy released upon relaxation of the velocities from variance $\propto T_i$ to $\propto T$. One could also say that the  overdamped literature \cite{Lapolla2020PRL,Dieball2023PRR,Ibanez2024NP} assumes velocity relaxation (occurring instantaneously at $t=0$) to occur \emph{before} the temperature quench (also happening instantaneously at $t=0$). The reason why we are able to detect this here is because we incorporate it in, and derive from, the more general underdamped (phase-space) framework.

Even though this may appear incomplete or problematic at first sight, the overdamped relaxation asymmetry and its proof \cite{Lapolla2020PRL,Ibanez2024NP} remain meaningful and valid for one particular reason: \emph{The choice of TE temperatures is the same for underdamped and overdamped dynamics} (in fact, it is the same for any OUP \cite{Lapolla2020PRL}), such that the instant relaxation of the velocities does \emph{not} change the excess free energy at later times. In other words, the overdamped comparison of heating and cooling neglecting velocities is valid on the appropriate time scales and the preceding velocity relaxation does not distort or corrupt the results.

\begin{figure}
    \centering
    \includegraphics[width=.9\linewidth]{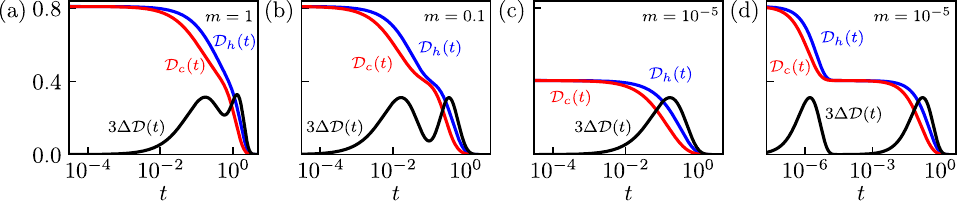}
    \caption{EFE for the process from Eq.~\eqref{underdamped SDE x 1d} with decreasing values of $m$ and $a=\gamma=k_BT=1$ and TE temperatures $k_BT_c=0.2$ and $k_BT_h=2.86$. Note that (c) and (d) show the same curves with a different left limit on the $x$-axis.}
    \label{fig:overdamped_Limit}
\end{figure}

This discussion is further illustrated in Fig.~\ref{fig:overdamped_Limit}, where we take $m/\gamma\to 0$. The different limits of the time axis in Fig.~\ref{fig:overdamped_Limit}c and d reflect two different overdamped limits. On the one hand, the velocity contributions are neglected and not observed since they occur at times that are not considered (corresponding to Fig.~\ref{fig:overdamped_Limit}c and, e.g., Ref.~\cite{Lapolla2020PRL}). On the other hand, they explicitly appear as long as we retain the underdamped description and also consider the shortest time-scales. Both considerations are contained in Theorem 1, but only the former is contained in the proofs in the overdamped literature \cite{Lapolla2020PRL,Dieball2023PRR,Ibanez2024NP}.
The important aspect that reconciles the two viewpoints is that the plateau in Fig.~\ref{fig:overdamped_Limit}d is at the same value for heating and cooling, which is (as mentioned before) due to the TE temperature condition being the same for all OUP. This reconciliation is lost for dynamics beyond OUP, where also the thermal relaxation asymmetry breaks down \cite{Lapolla2020PRL,Meibohm2021PRE}. 
Besides, note that such free energy considerations are relevant for the heat, entropy, and free energy in overdamped heat engines, since energy differences have to be considered \cite{Martinez2017SM} even though the dynamical equations may be overdamped (and thus do not explicitly include velocities).  

\section{Projected dynamics or marginalizations}\label{projections}
\begin{figure}
    \centering
    \includegraphics[width=.9\linewidth]{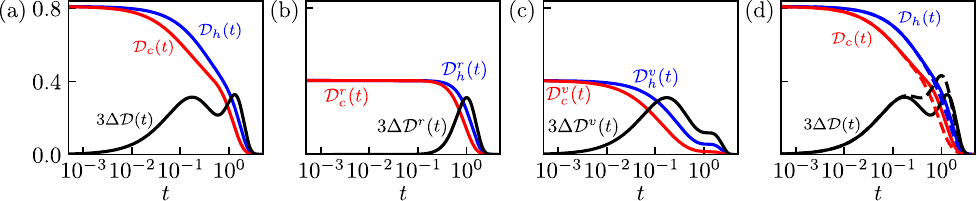}
    \caption{(a) EFE for the process from Eq.~\eqref{underdamped SDE x 1d} with $a=m=\gamma=k_BT=1$ and TE temperatures $k_BT_c=0.2$ and $k_BT_h=2.86$ alongside projections on the (b) $r$- and (c) $v$-component. (d) Comparison of the curves from (a) (solid lines in (d)) with the sum of the same-color lines in (b) and (c) (dashed lines in (d)) reveal that not all of the two-dimensional relaxation is contained in the one-dimensional marginalizations. The difference is due to off-diagonal terms in $\Sigma(t)$ that occur via Eq.~\eqref{solution dynamic Lyapunov}.}
    \label{fig:projection}
\end{figure}

In the overdamped limit of phase-space 
dynamics, we just compared the underdamped approach that also contains velocity with the overdamped description that only considers the position. Here, we present a complementary analysis by considering the marginalization or projection (these terms are used interchangeably) of underdamped dynamics on the position or velocity coordinate, individually. This corresponds to marginalizing the covariance matrix $\Sigma(t)$ on the position or velocity entry, respectively, on the diagonal and computing the excess free energy with respect to this marginal Gaussian distribution. 
For simplicity, we focus on the $1+1$-dimensional underdamped dynamics as in Eq.~\eqref{underdamped SDE x 1d} and show in Fig.~\ref{fig:projection} the EFE of marginalizations on $r$ and $v$, respectively. We see that the thermal relaxation asymmetry persists under this projection with the marginalized EFE $\Delta\mathcal D_{\rm marg}(t)$ defined as in Eq.~\eqref{EFE} but with the marginalized densities $\rho(x,t)\mapsto\rho_{\rm marg}(r,t)\equiv\int\rmd v\rho(x,t)$ when projecting on positions, or $\rho_{\rm marg}(v,t)\equiv\int\rmd r\rho(x,t)$ when projecting on velocities (and equivalently for $\rho_{\rm s}(x)$). In Sec.~\ref{sec:proof} we prove this extension of the thermal relaxation asymmetry that holds for underdamped OUP approaching equilibrium projected on position or velocity (or subspaces thereof). The precise statement is proven in Sec.~\ref{sec:proof_thm2} and reads as follows.
\medskip

\noindent \textit{\textbf{Theorem 2:} Assume underdamped stochastic dynamics Eq.~\eqref{underdamped SDE x} approaching an equilibrium state with block diagonal steady-state covariance 
$\Sigs = {\rm diag}(\Sigs^r,\Sigs^v)$. Consider a marginalization on the $d$-dimensional subspace reflecting $r$, or on the $d$-dimensional subspace reflecting $v$. Under such a marginalization, the difference of EFE of the resulting marginalized densities starting from TE temperatures fulfills the thermal relaxation asymmetry $\Delta\mathcal D_{\rm marg}(t)\ge 0$ for all $t\ge0$.}

\section{Outlook}\label{outlook}
We proved that the thermal relaxation asymmetry persists in the presence of inertial effects even for phase-space diffusion linearly driven into NESS, i.e., that far-from-equilibrium heating 
is faster than cooling also for underdamped diffusion. 
\blue{We thereby answered on outstanding theoretical question on the generality of the thermal relaxation asymmetry, which calls for experimental verification in the style of Ref.~\cite{Ibanez2024NP}.} 
The coupling of positions and velocities
emerging in the phase-space setting underscored, in a genuinely striking fashion, the intricate
dynamics of far-from-equilibrium thermal relaxation processes that do \emph{not} pass through local equilibria. In formulating the stochastic dynamics of relaxation into NESS we isolated the relevant relaxatory contribution to the entropy production, which has a weaker physical interpretation. The introduced phase-space representation may in turn prove useful in generalizing thermodynamics bounds (e.g., thermodynamic uncertainty relations 
\cite{Barato2015PRL,Dechant2018JPAMT,Dechant2021PRX,Koyuk2019PRL,Koyuk2020PRL,VanVu2023PRX,Dieball2023PRL} 
and correlation bounds \cite{Dechant2023PRL,Dieball2022JPA}) from the overdamped to the underdamped setting. 

We also investigated the overdamped limit, whereby the generalized approach highlighted that an excess free energy contribution from the velocity degrees of freedom in fact does
\emph{not} trivially vanish in the overdamped limit. Instead it is affected by the precise
interpretation of temperature quenches in overdamped systems, in particular of the notion of quenching the instantaneously-relaxing velocity degrees of freedom. In this respect, there are in fact two distinct notions of the overdamped limit of thermal relaxation. Generally there is thus a need to also include velocity degrees of freedom in free-energy considerations for overdamped dynamics whenever one faces non-constant (time-dependent) temperature (such as was done for Brownian heat engines in \cite{Martinez2017SM}).

Note that this was only the first attempt to analyze thermal relaxation in the full phase-space setting, which provokes many further intriguing questions. For example, the level-set geometry during thermal relaxation---rotations in opposite directions during heating and cooling in the overdamped setting \cite{Dieball2023PRR}---is particularly intriguing as it is unclear what happens close to critically coherent oscillations in the phase-space setting.
\blue{We expect that already the ($2$+$2$)-dimensional example in Eq.~\eqref{underdamped SDE x 2d} with $A_r$ generalized to elliptical potentials in the spirit of Ref.~\cite{Dieball2023PRR} will give rise to a rich phenomenology of level-set rotations motivating further studies.}
Moreover, a thermal kinematics framework \cite{Ibanez2024NP} in the phase-space setting remains elusive. 

\blue{In addition, note that viewed in in position space, underdamped dynamcis give rise to non-Markovian dynamics, i.e., memory effects. Therefore, this offers a route to address of thermal relaxation in non-Markovian dynamics, which could, e.g., be generalized to motion in harmonic potentials driven by correlated Gaussian noise.}

Furthermore, while we addressed projections onto position- and velocity-degrees of freedom, respectively, in the time-reversible setting, the asymmetry for projections in systems driven into NESS both, in presence and absence of inertia, remains an open question.  


\section{Proofs}\label{sec:proof}
Before we perform the proofs of Theorems 1 and 2, we first take a closer look at the setup, including the splitting of entropy production, and define further notation. 
In particular, we will perform a change of basis to facilitate separating symmetric and antisymmetric parts. 
Note that although the problem is defined in terms of stochastic dynamics, the proof is purely algebraic, owing to the linear structure of the OUP. 

\subsection{Non-adiabatic entropy production in underdamped dynamics}\label{sec:proof_underdamped_entropy}
For overdamped dynamics relaxing towards NESS, the Kullback-Leibler divergence $\mathcal D(t)$ corresponds to the \textit{non-adiabatic part of the entropy production} \cite{VandenBroeck2010PRE,Dieball2023PRR}. In contrast to dynamics relaxing towards equilibrium, this is no longer the full entropy production, and the interpretation as an EFE is lost (since free energies do not suffice to describe NESS). Nevertheless, the identification as the non-adiabatic entropy production justifies using $\mathcal D(t)$ as the appropriate observable to quantify thermal relaxation \cite{Dieball2023PRR}.
We here generalize the splitting of the entropy production highlighting the physical meaning of $\mathcal D(t)$ for relaxation towards NESS to underdamped dynamics.

For convenience, we here write the phase-space coordinate as $x=(q,p)$ with position $q=r$ and momentum $p=mv$ (the $q$ instead of $r$ is only used as a (redundant) reminder that we work with a different $x$ here). At least for the case of scalar $m$, all statements identically apply to $x=(r,v)$.

We start with the Langevin equation
\begin{align}
  dq_t&=m^{-1}p_tdt\nonumber\\
  dp_t&=F(q_t)dt-\gamma m^{-1}p_tdt+\sqrt{2k_{\rm B}T\gamma}dW_t
\end{align}
or, in terms of the phase-space vector $x=(q,p)^T\in\R^{2d}$
\begin{align}
  dx_t&=b(x_t)dt+\sqrt{2D_p}dW_t\label{phase space sde}
\end{align}
with drift vector and diffusion matrix (the index $p$ in $D_p$ is to avoid confusion of this matrix in $(q,p)$-space with the slightly different diffusion matrix in $(r,v)$-space)
\begin{align}
  b(x)= \begin{bmatrix} m^{-1}p \\  F(q)-\gamma m^{-1}p \end{bmatrix}, \quad  D_p\equiv\begin{bmatrix} 0 & 0\\ 0 & k_{\rm B}T\gamma   \end{bmatrix}\,.
\end{align}
The corresponding Klein-Kramers/Fokker-Planck equation then reads
\begin{align}
\partial_t\rhot&=[-\nabla \cdot  b(x)+ \nabla\cdot
  D_p\nabla]\rhot=-\nabla\cdot [b(x)-D_p\nabla]\rhot\nonumber\\
  &=-\nabla\cdot \gat\rhot
\end{align}
with $\nabla=(\nabla_q,\nabla_p)^T$. The local mean flow $\gat\in\R^{2d}$ is defined as (with $\nut$ as, e.g., in Ref.~\cite{Dieball2024PRL})
\begin{align}
    \gat&\equiv\frac{[b(x)-D_p\nabla]\rhot}{\rhot}\nonumber\\
    &=\begin{bmatrix} m^{-1}p \\  F(q)-\gamma [m^{-1}p +k_{\rm B}    T \nabla_p\ln \rhot] \end{bmatrix}\nonumber\\
    &=\begin{bmatrix} m^{-1}p \\  F(q)-\gamma \nut\end{bmatrix} \in\R^{2d}\nonumber\\
    \nut&\equiv m^{-1}p +k_{\rm B}T\nabla_p\ln \rhot \in\R^d\,,
\end{align}
We now generalize the splitting of the formula for the total entropy production into individual parts resembling the non-adiabatic and adiabatic parts as in Ref.~\cite{VandenBroeck2010PRE}. This assumes relaxation towards NESS, where $\gat$ and $\nut$ approach in the steady state time-independent vectors $\gas$ and $\nus$.

First note that
\begin{align}
\gat-\gas=\begin{bmatrix} 0 \\  -\gamma
(\nut-\nus)\end{bmatrix}=\begin{bmatrix} 0 \\  -k_{\rm B} T\gamma\nabla_p\ln\frac{\rhot}{\rhos}\end{bmatrix}=-D_p\nabla\ln\frac{\rhot}{\rhos}\,. \label{difference Gamma}
\end{align}
Since we want to obtain a splitting involving $\mathcal D(t)$ as in Eq.~\eqref{EFE}, consider the following calculations (in analogy to the overdamped calculations in \cite[Sec.~5.2.2]{Dieball_Thesis}),
\begin{align}
    \partial_t\int\rmd x&\rhot\ln\left[\frac{\rhot}{\rhos}\right]=\int\rmd x\ln\left[\frac{\rhot}{\rhos}\right]\partial_t\rhot\nonumber\\
    &=-\int\rmd x\ln\left[\frac{\rhot}{\rhos}\right]\nabla\cdot\gat\rhot\nonumber\\
    &=\Elr{\tgat\nabla\ln\left[\frac{\trhot}{\trhos}\right]}\nonumber\\
    &=\Elr{\begin{bmatrix} m^{-1}p_t \\  F(q_t)\end{bmatrix}\cdot\nabla\ln\left[\frac{\rhot}{\rhos}\right]}-\Elr{\begin{bmatrix} 0 \\ \gamma \nut\end{bmatrix}\cdot\nabla\ln\left[\frac{\trhot}{\trhos}\right]}\nonumber\\
    &\overset{\rm Eq.~\eqref{difference Gamma}}=\Elr{\begin{bmatrix} m^{-1}p_t \\  F(q_t)\end{bmatrix}\cdot\nabla\ln\left[\frac{\trhot}{\trhos}\right]}+\Elr{\tnut\gamma[\tnut-\tnus]}\label{calculation 1}
    \,.
\end{align}
Moreover, note that
\begin{align}
    \Elr{\tgas\cdot\nabla\ln\left[\frac{\trhot}{\trhos}\right]}&=\int\rmd x\gas\rhos\cdot\nabla\left[\frac{\rhot}{\rhos}\right]\nonumber\\
    &=-\int\rmd x\frac{\rhot}{\rhos}\nabla\cdot\gas\rhos=0\,,
\end{align}
since in the steady state $\nabla\cdot\gas\rhos=-\partial_t\rhos=0$. However, upon splitting as above we obtain
\begin{align}
    0&=\Elr{\tgas\cdot\nabla\ln\left[\frac{\trhot}{\trhos}\right]}\nonumber\\
    &=\Elr{\begin{bmatrix} m^{-1}p_t \\  F(q_t)\end{bmatrix}\cdot\nabla\ln\left[\frac{\trhot}{\trhos}\right]}+\Elr{\tnus\gamma[\tnut-\tnus]}\,.\label{calculation 2}
\end{align}
Subtracting Eq.~\eqref{calculation 2} from Eq.~\eqref{calculation 1}, we obtain similar to the overdamped result \cite{VandenBroeck2010PRE}
\begin{align}
    \partial_t\int\rmd x\rhot\ln\left[\frac{\rhot}{\rhos}\right]&=\Elr{[\tnut-\tnus]\gamma[\tnut-\tnus]}\,.\label{KL as square}
\end{align}
Note that also in analogy to the overdamped result we have for the total entropy production (see, e.g., Ref.~\cite{Dieball2024PRL})
\begin{align}
    \partial_t S_{\rm tot}=k_{\rm B}\Elr{\tnut\gamma\tnut}\,.
\end{align}
To obtain a splitting of his total entropy production rate, write using Eq.~\eqref{KL as square}
\begin{align}
    \partial_t S_{\rm tot}/k_{\rm B}&=\Elr{[\tnut-\tnus+\tnus]\gamma[\tnut-\tnus+\tnus]}\\
    &=\partial_t\int\rmd x\rhot\ln\left[\frac{\rhot}{\rhos}\right]+\Elr{\tnus\gamma\tnus}+2\Elr{\tnus\gamma[\tnut-\tnus]}\nonumber
    \,.
\end{align}
This would be in direct analogy to the overdamped result \cite{VandenBroeck2010PRE}, if $\Elr{\nus\gamma[\nut-\nus]}$ would be zero. However, this is generally not the case. To proceed, we use Eq.~\eqref{calculation 2} to note
\begin{align}
    \Elr{\tnus\gamma[\tnut-\tnus]}&=-\int\rmd x\rhot\begin{bmatrix} m^{-1}p \\  F(q)\end{bmatrix}\cdot\nabla\ln\left[\frac{\rhot}{\rhos}\right]\nonumber\\
    &=\int\rmd x\rhot\begin{bmatrix} m^{-1}p \\  F(q)\end{bmatrix}\cdot\nabla\ln\rhos\nonumber\\
    &=\Elr{\begin{bmatrix} m^{-1}p_t \\  F(q_t)\end{bmatrix}\cdot\nabla\ln\trhos}\,.\label{cross term}
\end{align}
where in the last line we used that the since vector commutes with $\nabla$ and $\rhot\nabla\ln\rhot=\nabla\rhot$, the first term in $\ln[\rhot/\rhos]=\ln\rhot-\ln\rhos$ vanishes after integration, see also Ref.~\cite[Eq.~(A4)]{Lyu2024PRE}. Note that the term in Eq.~\eqref{cross term} vanishes as $\rho(x,t)\to\rho_{\rm s}(x)$.
Overall, we write the rate of total entropy production
\begin{align}
    \partial_t S_{\rm tot}/k_{\rm B}&=\partial_t\Elr{\ln\left[\frac{\trhot}{\trhos}\right]}+\Elr{\tnus\gamma\tnus}+2\Elr{\begin{bmatrix} m^{-1}p_t \\  F(q_t)\end{bmatrix}\cdot\nabla\ln\trhos}\nonumber\\
    &=\frac{\partial_t\mathcal D(t)}{k_{\rm B}T}+\Elr{\tnus\gamma\tnus+2\begin{bmatrix} m^{-1}p_t \\  F(q_t)\end{bmatrix}\cdot\nabla\ln\trhos}\,.
\end{align}
In the overdamped case, the $\mathcal D(t)$-part of the entropy production was called \textit{non-adiabatic} part of the entropy production and $\Elr{\nus\gamma\nus}$ was called the \textit{adiabatic} part. Here, a third term appears. Since this third term, similar to the adiabatic part, only involves $\rhot$ via the average $\E{\cdots}$, we decide to attribute this term to the adiabatic (``housekeeping'' \cite{VandenBroeck2010PRE,Seifert2012RPP}) part of the entropy production and (as in the overdamped case \cite{Dieball2023PRR}) only consider the $\mathcal D(t)$-part for the interpretation of relaxation. 

\subsection{Details on the positivity of real parts of eigenvalues of $A$}\label{sec:proof_positive_real_parts}
Although the proof of Theorem 1 does not hinge on the specific form of $A$ for underdamped dynamics, we first provide a proof that for any $A_r$ that is diagonalize with positive (real) eigenvalues and scalar $m,\gamma>0$, we have that the resulting matrix $A$ for underdamped dynamics as in Eq.~\eqref{underdamped SDE x} (with no velocity-dependent forces beyond Stokes friction, i.e., $A_v=0$) has exclusively eigenvalues with positive real parts. This proof is not strictly necessary for proof of the Theorems, but it nicely shows that (as expected) the class of processes that fulfill the assumptions of the theorem is sufficiently large to make physically relevant statements. To prove this, assume $GA_rG^{-1}/m={\rm diag}(a_1,\dots,a_d)$ with $s_i>0$, and note that using block matrices
\begin{align}
    \begin{bmatrix} 0&G\\G&0\end{bmatrix}A\begin{bmatrix} 0&G\\G&0\end{bmatrix}^{-1} 
    &=\begin{bmatrix} 0&G\\G&0\end{bmatrix}\begin{bmatrix}0&-1\\A_r/m&\gamma/m\end{bmatrix}\begin{bmatrix} 0&G^{-1}\\G^{-1}&0\end{bmatrix}\nonumber\\
    &= \begin{bmatrix} 0&G\\G&0\end{bmatrix}\begin{bmatrix} -G^{-1}&0\\\gamma G^{-1}/m&A_rG^{-1}/m\end{bmatrix}\nonumber\\
    &=\begin{bmatrix} \gamma/m&{\rm diag}(a_i)\\-1&0\end{bmatrix}\,.
\end{align}
Let $P=(P^T)^{-1}$ be the permutation corresponding to reordering $1,2,\dots,2d$ to $1,d+1,2,d+2,\dots,d,2d$ to obtain
\begin{align}
    P\begin{bmatrix} 0&G\\G&0\end{bmatrix}A\begin{bmatrix} 0&G\\G&0\end{bmatrix}^{-1}P^{-1} &= \begin{bmatrix}
        {\gamma/m}&{a_1}&0&0&\cdots&\cdots&0&0\\
        {-1}&{0}&0&0&\cdots&\cdots&0&0\\
        0&0&{\gamma/m}&{a_2}&\cdots&\cdots&0&0\\
        0&0&{-1}&{0}&\cdots&\cdots&0&0\\
        \cdots&\cdots&\cdots&\cdots&\cdots&\cdots&\cdots&\cdots\\
        \cdots&\cdots&\cdots&\cdots&\cdots&\cdots&\cdots&\cdots\\
        0&0&0&0&\cdots&\cdots&{\gamma/m}&{a_d}\\
        0&0&0&0&\cdots&\cdots&{-1}&{0}
    \end{bmatrix}\label{block form}
\end{align}
Then, in each $2\times2$ block we have eigenvalues $\lambda_1\lambda_2=\det({\rm Block}_i)=a_i>0$ and $\tr({\rm Block}_i)=\gamma/m=\lambda_1+\lambda_2$. If the eigenvalues $\lambda_{1,2}$ are real, then both are positive since their product and sum is positive. If they are not real, then we have a pair of complex conjugate eigenvalues such that real part of both is $\gamma/2m>0$. Finally, note that $A$ has the same eigenvalues as the block diagonal matrix in Eq.~\eqref{block form} since they only differ by a change of basis, and note that the eigenvalues of the block diagonal matrix are given by the eigenvalues of the $d$ individual $2\times2$ blocks.

We have thus shown that, for $A_r$ diagonalize with positive (real) eigenvalues and scalar $m,\gamma>0$, the eigenvalues of $A$ may be complex or not, but have always positive real parts (as required to approach steady-state dynamics and to prove the theorems). Note that $A_r$ being diagonalize with positive (real) eigenvalues does not only include dynamics approaching equilibrium, but for asymmetric $A_r\ne A_r^T$ can also reflect moderately driven dynamics \cite{Dieball2023PRR}.
For complex eigenvalues of $A_r$, the eigenvalues of $A$ may still all have positive real parts (see example in Eq.~\eqref{underdamped SDE x 2d} and Fig.~\ref{fig:trajectories_and_EFE}c,d), but one may also have the case where eigenvalues of $A$ may have negative real parts (reflecting coherent oscillations that overcome the confinement) even though those of $A_r$ only have positive real parts (e.g., for the process in Eq.~\eqref{underdamped SDE x 2d} with $a=m=\gamma=1$ and $\abs{\omega}>1$).

\subsection{Proof of Theorem 1}\label{sec:proof_thm1}
\subsubsection{$\Delta\mathcal{D}(t)$ in terms of the covariance}\ \\
Since the OUP in Eq.~\eqref{SDE in theorem} is a Gaussian process, the excess free energy $\mathcal{D}(t)$ in Eq.~\eqref{EFE} can be expressed in terms of the mean and covariance of the time-dependent density $\rho(x,t)$ and steady-state density $\rho_{\rm s}(x)$. In the framework of the thermal relaxation asymmetry, the initial condition is a zero-mean Gaussian with covaraince $\Sigma_{h,c}(0)=T_{h,c}\Sigs/T$, and according to the time-evolution in Eq.~\eqref{solution dynamic Lyapunov}, the mean remains zero while the covariance obeys 
\begin{align}
    X(t) &\equiv \rme^{-At}\Sigs\rme^{-A^Tt}\Sigs^{-1}\nonumber\\
    \dThc &\equiv \frac{T_{h,c}}T-1\nonumber\\
    \Sigma_{h,c}(t)\Sigs^{-1} &=\mathbbm 1+\dThc X(t)\,.\label{def X}
\end{align}
where $\mathbbm 1$ denotes the unit matrix. 
The Kullback-Leibler divergence of multivariate zero-mean Gaussian densities $P_{1,2}$ reads $2D_{\rm KL}\left(P_1||P_2\right)=-\ln\left(\det\left[\Sigma_1\Sigma_2^{-1}\right]\right)+\tr\left(\Sigma_1\Sigma_2^{-1}-\mathbbm 1\right)$, which for $\Sigma_1=\Sigma_{h,c}(t)$ and $\Sigma_2=\Sigs$ gives \cite{Dieball2023PRR}
\begin{align}
    \mathcal D_{h,c}(t) &= \frac{1}{2} \dThc\,\tr  X(t)-\frac{1}{2}\ln\det\left[\mathbbm 1+ \dThc\, X(t)\right], \nonumber\\
\Delta\mathcal D(t) &\equiv \mathcal D_h(t)-\mathcal D_c(t) = \frac{ \dT_h- \dT_c}{2}\,\tr  X(t)-\frac{1}{2}\ln\frac{\det\left[\mathbbm 1+ \dT_h\, X(t)\right]}{\det\left[\mathbbm 1+ \dT_c\, X(t) \right]}\,. \label{expression DKL}
\end{align}
This phrases the question of proving the thermal relaxation asymmetry $\Delta\mathcal D(t)>0$ for TE temperatures $T_{h,c}$ into an algebraic problem. The TE initial conditions are exactly as in the overdamped case \cite{Lapolla2020PRL} characterized by $\mathcal D_h(0)=\mathcal D_c(0)$, i.e., since $ X(0)=\mathbbm 1$ the condition for TE temperatures is $\dT_h-\ln(1+\dT_h)=\dT_c-\ln(1+\dT_c)$.

\subsubsection{Sketch of the proof}\ \\
We here first sketch the proof of Theorem 1, i.e., the proof of $\Delta\mathcal D(t)>0$ for Eq.~\eqref{expression DKL}. 
This mainly builds up on the matrix formulation that was also used for overdamped driven OUP \cite{Dieball2023PRR}. It turns out that generalization to driven dynamics is very similar to the underdamped problem, since in both cases the challenge is that due to asymmetric drift matrices, the dynamics can no longer be reduced to one-dimensional systems (as is possible for overdamped OUP without driving \cite{Lapolla2020PRL}).

We therefore follow the steps (and notation) of the thermal relaxation asymmetry for driven dynamics from Ref.~\cite{Dieball2023PRR}. First note that the difference in EFE from Eq.~\eqref{expression DKL} can be reformulated in terms of the eigenvalues $x_{j}$, $j=1,2,\dots,2d$ of the matrix $X(t)$ as
\begin{align}
    \Delta\mathcal D(t) &= \sum_{j=1}^{2d}\nonumber\Delta\mathcal D(t;j)\\
    \Delta\mathcal D(t;j) &\equiv\frac{ \dT_h- \dT_c}{2}x^t_j-\frac{1}{2}\ln\left[\frac{1+ \dT_hx^t_j}{1+ \dT_cx_j^t}\right]\,.\label{expression DKL eigenvalues}
\end{align}
To conclude the relaxation asymmetry in driven \textit{overdamped} dynamics \cite{Dieball2023PRR}, it sufficed to show that $x_j^t\in(0,1)$ for all $j$ and $t>0$ since then for each $j$ the difference is positive, i.e., $\Delta\mathcal D(t;j)>0$. For \textit{underdamped} dynamics (generally speaking, for positive semi-definite $D$ instead of positive definite $D$), we have to slightly weaken this strategy and will instead show that $x_j^t\in(0,1]$ for all $j$ and $t>0$ and that for any $t>0$ there is at least one $j$ for which $x_j^t<1$. This also suffices to show $\Delta\mathcal D(t)>0$ for all $t>0$ since $x_j^t\in(0,1]$ implies $\Delta\mathcal D(t;j)\ge 0$ (equality for $x_j^t=1$ holds by definition of TE temperature). 

\subsubsection{Preparation of the proof}\label{sec:proof_preparation}\ \\
To prepare the proof, we introduce some notation and a useful change of basis. Since $\Sigs^{-1}$ is symmetric with positive eigenvalues by the assumptions of Theorem 1, we can diagonalize it as $ O\Sigs^{-1} O^T={\rm diag}(s_j)$ with $O^{-1}=O^T$ and define $ B\equiv  O{\rm diag}(\sqrt{s_j}) O^T$ such that we have a matrix $B$ with $B= B^T$ and $B^2=\Sigs^{-1}$. 
Moreover, we introduce a rewriting that splits $A$ into contributions of the symmetric $D$ and an antisymmetric $\alpha$ as
\begin{align}
    A &=( D+\alpha)\Sigs^{-1}\nonumber\\
    \alpha&\equiv A\Sigs -  D\nonumber\\
    \alpha+\alpha^T &=  A\Sigs -  D+\Sigs A^T -  D\overset{\rm Eq.~\eqref{static Lyapunov}}=0\,. \label{definition alpha}
\end{align}
Then, we perform a change of basis using $B$ to obtain
\begin{align}
    \widetilde{ A}\equiv BAB^{-1}= B( D+\alpha) B\,.\label{A tilde}
\end{align}
Note that $ B D B= B\sigma( B\sigma)^T$ is positive semi-definite since, as shown before, every matrix of the form $MM^T$ (for $M$ real) is positive semi-definite. Also note that $B\alpha B=-(B\alpha B)^T$ is antisymmetric due to $\alpha^T=-\alpha$, see Eq.~\eqref{definition alpha}.
Further define
\begin{align}
    \widetilde{ X}(t) &\equiv  B X(t) B^{-1}
    \overset{\rm Eq.~\eqref{def X}}= B\rme^{- At} B^{-2}\rme^{- A^Tt} B^2 B^{-1}
    = \rme^{-\widetilde{ A}t}\left(\rme^{-\widetilde{ A}t}\right)^T\,,\label{X tilde}
\end{align}
with $X(t)$ from Eq.~\eqref{def X}. 

\subsubsection{Positivity of the eigenvalues}
We here show $x_j^t>0$ by first showing $x_j^t\ge0$ and then $x_j^t\ne0$. To obtain the former, use Eq.~\eqref{X tilde} to see that we have $\widetilde{ X}(t)=MM^T$ for some real matrix $M$ which already implies positive semi-definite, i.e., since $X(t)$ and $\widetilde{X}$ have the same eigenvalues (the matrices only differ by a change of basis), we have $x_j^t\ge 0$.

To prove that $x_j^t\ne0$, consider
\begin{align}
    \prod_{j=1}^{2d} x_j^t=\det[ X(t)]=\det\left(\rme^{-\widetilde{ A}t}\right)\det\left(\rme^{-\widetilde{ A}^Tt}\right)=\rme^{-2\tr(\widetilde{ A})t}\ne 0\quad \Rightarrow\quad  \forall j\colon x_j^t\ne0\,.
\end{align}
Thus we have $x_j^t>0$ for all $j$ and $t\ge 0$.

\subsubsection{Eigenvalues bounded by $1$}
To show that $x_j^t\le 1$ for all $j$ and $t\ge 0$, we again follow Ref.~\cite{Dieball2023PRR} and employ the log-norm inequality \cite{Dahlquist1958} with matrix norm/operator norm $\norm{\cdot}$, Euclidean vector norm $\norm{\cdot}_2$ and log norm $\mu(\cdot)$, 
\begin{align}
    \norm{N}&\equiv\sup_{v\in\R^{2d}\setminus 0}\frac{\norm{Nv}_2}{\norm{v}_2}\nonumber\\
    \mu( M)&\equiv\lim_{h\to0^+}\frac{{\norm{\mathbbm 1+h M}-1}}h\nonumber\\
    \norm{\exp( Mt)}&\le\rme^{\mu( M)t}
    \,.\label{log norm inequality}
\end{align}
The log norm $\mu(M)$ is given by the largest eigenvalue of the symmetric part of $M$, see \cite[Sec.~IX in Supplmental Material]{Dieball2023PRR}. Therefore, $\mu(-\widetilde{A}^T)=\mu(-\widetilde{A})=\mu(-BDB)\le0$, since $BDB$ is positive semi-definite. Via the log-norm inequality in Eq.~\eqref{log norm inequality} this gives (by submultiplicativity of the matrix norm)
\begin{align}
    \norm{\widetilde{X}}\le\norm{\rme^{-\widetilde{A}t}}\norm{\rme^{-\widetilde{A}^Tt}}\le\rme^{\mu(-\widetilde{A})t}\rme^{\mu(-\widetilde{A}^T)t}\le 1\,.
\end{align}
Therefore, the (positive) eigenvalues of $X(t)$ are bounded by $x_j^t\le||\widetilde{X}||\le1$.

\subsubsection{One eigenvalue is strictly less than $1$}
Recall that $BDB$ is semi-definite but $BDB\ne 0$, i.e., all eigenvalues are non-negative and $\tr(BDB)>0$. Note that $BDB=0$ is excluded by Eq.~\eqref{static Lyapunov} since we assume positive definite $\Sigs$.
Since $\tr(B\alpha B)=0$, this implies $\tr(\widetilde{A})=\tr(BDB)+\tr(B\alpha B)>0$ and via Eq.~\eqref{X tilde}, $\prod_{j=1}^{2d}x_j^t=\det(\widetilde{ X})=\rme^{-2\tr(\widetilde{ A}t)}<1$ for $t>0$. Since all $x_j^t\in(0,1]$, this implies that for any $t>0$, there is at least one $j^*$ such that $x_{j^*}^t<1$.
This completes the proof of Theorem 1, i.e., in particular the thermal relaxation asymmetry persists to underdamped OUP. \hfill$\square$

\subsection{Proof of Theorem 2}\label{sec:proof_thm2}

\noindent We now prove Theorem 2, i.e., that the thermal relaxation asymmetry persists under projection/marginalization on the position or velocity coordinates, respectively, for underdamped dynamics approaching equilibrium. We assumed in Theorem 2 a block diagonal $\Sigs$,

\begin{align}
    \Sigs &= \begin{bmatrix} \Sigs^r &0\\0&\Sigs^v \end{bmatrix}\nonumber\\
    \Sigs^{-1} &= \begin{bmatrix} (\Sigs^r)^{-1} &0\\0&(\Sigs^v)^{-1} \end{bmatrix}\,,\label{Sigs block diagonal}
\end{align}
which typically reflects the Maxwell-Boltzmann equilibrium state in a harmonic potential $\phi(x)=\phi_r(r)+\phi_v(v)$.
Note that Theorem 2 would be obvious for block diagonal $\Sigma_{h,c}(t)$, but this is not given, see Eq.~\eqref{solution dynamic Lyapunov}. 
As a consquence of Eq.~\eqref{Sigs block diagonal}, the change of basis $B=\sqrt{\Sigs^{-1}}$ as defined in Sec.~\ref{sec:proof_preparation} also takes on a block diagonal form
\begin{align}
    B=\begin{bmatrix} B^r &0\\0&B^v \end{bmatrix}\,. \label{B block diagonal}
\end{align}
Defining $\mathcal P$ to be the operator projecting onto the subspace that we marginalize on [e.g., for the position-marginals we would have $\mathcal P(\Sigs)=\Sigs^r$], the EFE difference $\Delta\mathcal D(t)$ of the marginalized process reads
\begin{align}
     X_{\rm marg}(t) &\equiv  \mathcal P\left(\rme^{- A t}\Sigma_{{\rm s},w}\rme^{- A^T t}\right)\left[\mathcal P(\Sigma_{{\rm s},w})\right]^{-1},\nonumber\\
    \Delta\mathcal D_{\rm marg}(t) &= \frac{ \dT_h- \dT_c}{2}\,\tr  X_{\rm marg}(t)-\frac{1}{2}\ln\frac{\det\left[\mathbbm 1+ \dT_h\, X_{\rm marg}(t)\right]}{\det\left[\mathbbm 1+ \dT_c\, X_{\rm marg}(t) \right]}\label{delta D marginal}
\end{align}
As before, the proof reduces to proving that the eigenvalues of $ X_{\rm marg}(t)$ fulfill $x^j_{\rm marg}\in(0,1]$. Writing the relevant matrices in terms of $d\times d$ blocks, we have for the projection on the spatial coordinates
\begin{align}
    \rme^{- A t}\Sigma_{{\rm s},w}\rme^{- A^T t} &\equiv\begin{bmatrix}M_{rr}(t)&M_{rv}(t)\\M_{vr}(t)&M_{vv}(t)\end{bmatrix}\nonumber\\
    X_{\rm marg}(t)&=M_{rr}(t)(\Sigs^r)^{-1} \nonumber\\
    \mathcal P[\widetilde{X}(t)]&=\mathcal P\left(\begin{bmatrix} B^r &0\\0&B^v \end{bmatrix}\begin{bmatrix}M_{rr}(t)&M_{rv}(t)\\M_{vr}(t)&M_{vv}(t)\end{bmatrix}\begin{bmatrix}(\Sigs^p)^{-1} &0\\0&(\Sigs^{\neg p})^{-1}\end{bmatrix}\begin{bmatrix} B^r &0\\0&B^v \end{bmatrix}^{-1}\right)\nonumber\\
    &=B^rM_{rr}(t)(\Sigs^p)^{-1}(B^r)^{-1}\nonumber\\
    &=B^rX_{\rm marg}(t)(B^r)^{-1}\,.
\end{align}
The projection on the velocity coordinates is completely analogous with swapped indices $r\leftrightarrow v$. 
Since $B^r$ represents again only a change of basis, we conclude that $X_{\rm marg}(t)$ has the same eigenvalues as $\mathcal P[\widetilde{X}(t)]$. Recall that we found that $\widetilde{X}(t)$ is symmetric with eigenvalues $x^j\equiv x_j^t\in(0,1]$, see Sec.~\ref{sec:proof_thm1}.
This directly implies that the eigenvalues of $\mathcal P[\widetilde{X}(t)]$ fulfill $x_{\rm marg}^j\in(0,1]$, which actually concludes the proof of Theorem 2.
To show this implication from the eigenvalues of $\widetilde{X}(t)$ to the ones of $\mathcal P[\widetilde{X}(t)]$, we perform a proof by contradiction. 

To this end, assume there was an eigenvalue ${x^j_{\rm marg}\not\in(0,1]}$ with normalized eigenvector $v_{\rm marg}$ such that ${v_{\rm marg}^T\mathcal P[\widetilde{X}(t)]v_{\rm marg}=x^j_{\rm marg}\not\in(0,1]}$.~Take the vector $v\in\R^{2d}$ that has zeros in all entries lost in the projection and $\mathcal P(v)=v_{\rm marg}$.~Then $v$ is normalized and due to the zero entries we have $v^T\widetilde{X}(t)v=v_{\rm marg}^T\mathcal P[\widetilde{X}(t)]v_{\rm marg}=x^j_{\rm marg}\not\in(0,1]$. However, this is impossible since it contradicts the min-max-principle \cite{Horn_1985} which states for symmetric $\widetilde{X}(t)$ with eigenvalues $x^j\in(0,1]$ that $0<\min_jx^j\le v^T\widetilde{X}(t)v\le\max_jx^j\le 1$. Note that we are content with proving $x_{\rm marg}^j\in(0,1]$ instead of $x_{\rm marg}^j\in(0,1)$ since we allowed for equality $\Delta\mathcal D_{\rm marg}(t)\ge 0$ in the statement of Theorem 2.\hfill$\square$

\section*{Acknowledgments}
Financial support from the European Research Council (ERC) under the European Union’s Horizon Europe research and innovation program (grant agreement No 101086182 to AG) and 
as well as the German Research Foundation
(DFG) through the Heisenberg Program grant
GO 2762/4-1 project number 519908342 to AG)
is gratefully acknowledged.

\section*{References}
\bibliography{underdamped_relaxation}
\end{document}